\documentclass[prd,twocolumn,superscriptaddress,altaffilletter,showpacs,nofootinbib]{revtex4}

\usepackage{amssymb}
\usepackage[dvips]{graphicx}
\usepackage{amsmath}

\newcommand{\be}{\begin{equation}}
\newcommand{\ee}{\end{equation}}
\newcommand{\bea}{\begin{eqnarray}}
\newcommand{\eea}{\end{eqnarray}}

\newcommand{\der}{\partial}
\newcommand{\vphi}{\varphi}

\begin{document}


\title{On a symmetry relating gravity with antigravity}

\author{Israel Quiros}

\email{iquiros@fisica.ugto.mx}

\affiliation{Departamento de Ingenier\'ia Civil, Divisi\'on de Ingenier\'ia, Universidad de Guanajuato, Guanajuato, CP 36000, M\'exico}

\date{\today}

\begin{abstract}
I investigate the impact of a ``would be'' fundamental symmetry of the laws of nature under the interchange of gravity and antigravity, on the understanding of negative energies in general relativity. For this purpose a toy model that is based on Einstein-Hilbert gravity with two minimally coupled self-interacting scalar fields is explored, where the second (exotic) scalar field with negative energy density may be regarded, alternatively, as an antigravitating field with positive energy. Spontaneous breakdown of reflection symmetry is then considered in order to discuss the implications the proposed ``would be'' fundamental symmetry might have for the vanishing of the cosmological constant. A possible connection of the gravity-antigravity symmetry with the so called quintom field is also explored.\end{abstract}


\maketitle


\section{Introduction}\label{intro}

One of the most profound mysteries in fundamental physics is the cosmological constant problem (see \cite{ccp-weinberg, ccp-padma, ccp-carroll, ccp-sahni, ccp-vilenkin, ccp-rubakov, ccp-dolgov, ccp-other} for well-known reviews on this subject). The physical basis for the cosmological constant $\Lambda$ are the zero-point vacuum fluctuations. The expectation value of the energy-momentum tensor for vacuum can be written in the Lorentz invariant form $\left\langle T_{\mu\nu}\right\rangle_\text{vac}=(\Lambda/8\pi G)g_{\mu\nu}$, where $G$ is the Newton's constant.\footnote{In the natural units adopted in this paper, where the speed of light $c=\hbar=1$, the Planck mass is related with the Newton's constant through the known relationship $M^2_\text{Pl}=1/8\pi G$.} It is divergent both for bosons and for fermions. Since bosons and fermions (of identical mass) contribute equally but with opposite sign to the vacuum expectation of physical quantities, supersymmetry was expected to account for the (nearly) zero value of the cosmological constant, through an accurate balance between bosons and fermions in nature. However, among other objections, the resulting scenario is not the one it is expected to occur if a universe with an early period of inflation (large $\Lambda$) and a very small current value of $\Lambda$ is to be described \cite{ccp-sahni} and, besides, supersymmetry is badly broken in nature at high energies $\sim 10^2$ GeV \cite{ccp-padma}. This leads to an unacceptable large value of the vaccum energy density $\rho_\text{vac}$. Although other mechanisms and principles -- among them a running $\Lambda$ and the anthropic principle -- have been invoked to solve the cosmological constant (vacuum energy density) puzzle, none of them have been able to give a definitive answer to this question and the problem still remains a mystery \cite{ccp-weinberg, ccp-padma, ccp-carroll, ccp-sahni, ccp-vilenkin, ccp-rubakov, ccp-dolgov, ccp-other}. A related (also unresolved) question is that, besides supersymmetry, there is no other known fundamental symmetry in nature which will set to zero the value of $\Lambda$. Hence, the search for a feasible candidate of such a symmetry also remains a challenge. 

In the present paper I want to further develop an earlier proposal, where the possibility that such a fundamental symmetry could be the one generated by the interchange of gravity and antigravity, was explored for the first time \cite{quiros-2004}. The intuitive idea behind this possibility is that, once this symmetry is incorporated into the theory of gravitation coupled to the standard model of particles (SMP), to each gravitating (G) particle/antiparticle, it corresponds an antigravitating (aG) partner, whose contribution to the vacuum energy exactly cancels the gravitating particle's contribution. As it is for supersymmetry, the adoption of the gravity-antigravity (G-aG) symmetry would double the number of existing standard model particles and antiparticles \cite{gag-sabine}. While in the absence of gravity -- as, for instance, in Minkowski vacuum -- the G-particles/antiparticles and their aG-partners are the same, as long as gravity is switched on, these are differentiated by their interactions with the gravitational field.\footnote{The relative strength of the expected effect of switching on gravity, when compared with the strength $\alpha\approx 1/137$ of the electromagnetic interactions, is of the order $\sim 10^{-36}$. Hence, one should not expect to discover antigravitating particles/antiparticles in standard experiments with a cloud or a bubble-chamber.} Also, as a result of the G-aG symmetry, the only allowed vacuum processes are those which satisfy the conservation of the gravitational charge. For instance, virtual particles created out of the vacuum can arise in pairs of the form (G-particle, aG-antiparticle), or (aG-particle, G-antiparticle), but never in pairs like (G-particle, G-antiparticle), etc. Naively the whole picture can be visualized as if there were two distinct vacua: one gravitating and other one that antigravitates so that, assuming an exact balance -- independent of the mean energy of the gravitating/antigravitating vacua -- the resulting net (averaged) vacuum does not gravitate at all. 

The kind of symmetry I am proposing to account for the (nearly) zero value of the cosmological constant, opens up the possibility that such exotic entities like aG-objects, might exist in nature. So, why this kind of objects are not being observed in our universe? A similar question has been raised before in the context of matter-antimatter symmetry \cite{sakharov, m-am, bar-asym, bar-asym-1, bas-cdm, pap-asym}. In this last case a possible mechanism for generating the desired amount of baryon asymmetry relies on three necessary (Sakharov's) conditions \cite{sakharov}: i) baryon number non-conservation, ii) C and CP violation and iii) deviations from thermal equilibrium. In the same fashion, an answer to the problem of gravitating-antigravitating matter asymmetry could be approached. In this sense, if accept a nonvanishing net vacuum energy density, one should expect an (perhaps very tiny) amount of violation of conservation of the gravitational charge associated with G-aG symmetry. 

A very encouraging aspect of the symmetry between gravity and antigravity is its ability to deal with negative energies. As we shall see, if take for serious the G-aG symmetry, the occurrence of negative energies in general relativity is not as harmful as thought: the negative energy gravitating fields may be regarded as positive energy fields which antigravitate.

Although the approach undertaken in this paper is fully classical, consideration of quantum effects is of prime importance to understand the actual feasibility of G-aG symmetry to occur in nature. This is due to the potential danger of dealing with negative energies even if these can be viewed as positive through trading gravity by antigravity. Anyway, the idea of adopting G-aG symmetry as a fundamental symmetry of nature is so exciting that even a classical approach to it seems interesting enough.


\section{G-aG symmetry}\label{gag} 

Before pushing the G-aG idea any further, let us note that the G-aG transformation: $G\rightarrow-G$, may be viewed alternatively as the interchange of the Planck mass squared and the negative (tachyonic) Planck mass squared: $$G\rightarrow -G\;\Leftrightarrow\;M_\text{Pl}^2\rightarrow -M_\text{Pl}^2.$$ Hence, simultaneously with the interchange of gravity and antigravity, signature reversal: $g_{\mu\nu}\rightarrow-g_{\mu\nu}$ \cite{duff}, is also required in order to preserve causality. Actually, let us consider the Klein-Gordon equation for a massive field $\psi$:\footnote{In this paper I chose the East coast signature for the metric: ($-+++$).} $$\left(\der^2-m^2\right)\psi=0,$$ where $\der^2\equiv g^{\mu\nu}\der_\mu\der_\nu$. Even without making statements on the G-aG symmetry we see that, under the replacement of a particle by its tachyonic partner: $m^2\rightarrow-m^2$ ($m\rightarrow\pm im$), signature reversal $g_{\mu\nu}\rightarrow-g_{\mu\nu}$ $\Rightarrow\;\der^2\rightarrow-\der^2$ is simultaneously required in order to preserve the equation of motion. 

In this paper I shall explore the symmetry relating gravity with antigravity by investigating classical actions of the following form:

\bea S=\int d^4x\sqrt{|g|}\left[\frac{R}{16\pi G}+{\cal L}(\psi,\der\psi)-{\cal L}(\bar\psi,\der\bar\psi)\right],\label{gag-action}\eea where ${\cal L}(\psi,\der\psi)$ is the Lagrangian for the gravitating fields $\psi$, while ${\cal L}(\bar\psi,\der\bar\psi)$ is the Lagrangian for their antigravitating partners $\bar\psi$. Notice that the full Lagrangian in (\ref{gag-action}) times $16\pi G$, can be written in the following way: $R+16\pi\left[G{\cal L}(\psi,\der\psi)+(-G){\cal L}(\bar\psi,\der\bar\psi)\right]$, where it is made explicit that the unconventional negative sign of the ghost Lagrangian ${\cal L}(\bar\psi,\der\bar\psi)$, can be absorbed into the Newton's constant, so as to make manifest that the field $\bar\psi$ carries conventional energy signature at the cost of antigravitating. 

The action (\ref{gag-action}) has been explored before in \cite{gag-linde} to address the cosmological constant problem but it has not been related with G-aG symmetry in that reference.\footnote{In recent years the model of \cite{gag-linde} has been exploited also in the context of the so called ``energy-parity'' in scenarios with a ghost sector which complements the ``visible matter'' sector \cite{sundrum, elze}.} It is straightforward noting that, the pure gravity part of (\ref{gag-action}): $S_g=\int d^4x\sqrt{|g|} R/(16\pi G)$, is by itself G-aG symmetric. Actually, under signature flip: $g_{\mu\nu}\rightarrow-g_{\mu\nu}$, the Ricci tensor is unchanged $R_{\mu\nu}\rightarrow R_{\mu\nu}$, while the curvature scalar $R\rightarrow -R$, so that, as long as simultaneously $G\rightarrow-G$, the ratio $R/G$ is unaltered. This means that, whenever gravitating and antigravitating sectors are decoupled, gravity is attractive as felt by both observers living in the G sector and in the aG one. However, in the case when the G and aG sectors are mixed together, gravity, assuming it is generated by a G-matter source, is attractive for the G-particles while it is repulsive for the aG-particles. The contrary situation is true if the gravitational field is generated by an aG-matter source. 

Now I will discuss on how harmless the negative energies could be if adopt the G-aG symmetry as a fundamental principle of physics.\footnote{To learn about the recurrence of negative energy densities in physics, see reference \cite{negenerg, sawicki, vilenkin}.} Let us assume an isotropic and homogeneous matter distribution within the three-volume $\Omega$, with energy density $\rho$, so that the gravitational mass inside that volume is given by: $M=\int_\Omega d^3x\,\rho$. If $\rho$ were a negative quantity ($\rho<0$), so were the gravitational mass $M<0$. The Newtonian gravitational potential of that matter distribution is given by: $\Phi=-GM/r$ ($r\equiv|{\bf r}|$). This is positive since we are assuming that $M<0$, hence, any test particle with positive gravitational mass $m$ would feel gravitational repulsion of strength $\vec{f}=-m\vec{\der}\Phi$, from the mass $M$. A more plausible physical interpretation is possible if assume G-aG symmetry to be a fundamental symmetry of physics. In this latter case, instead of a repulsive gravitational field generated by a negative gravitational mass ($M<0$) acting on the test particle $m$, one adopts the alternative picture where a repulsive antigravitational field ($G<0$) is generated by a positive mass $M>0$, i. e., by a positive energy density $\rho>0$.


\section{Fermions and G-aG symmetry}\label{fermions}

The action (\ref{gag-action}) is G-aG symmetric either, if the Lagrangian of the G-matter fields (and of their aG-partners) is unchanged by the G-aG transformations: $G\rightarrow-G$, $g_{\mu\nu}\rightarrow-g_{\mu\nu}$, or, if the following transformation is satisfied simultaneously with the G-aG transformations above: $${\cal L}(\chi,\der\chi)\rightarrow-{\cal L}(\chi,\der\chi),\;{\cal L}(\bar\chi,\der\bar\chi)\rightarrow-{\cal L}(\bar\chi,\der\bar\chi),\;\chi\rightarrow\bar\chi.$$ 

In order to explore this issue, let us consider a spin $1/2$ Dirac spinor field $\psi$, where I assume the Dirac gamma matrices to obey $\{\gamma_\mu,\gamma_\mu\}=2g_{\mu\nu}{\bf 1}$. Hence, since I chose the East coast signature $(-+++)$, the Dirac Lagrangian reads:

\bea {\cal L}_\psi=\frac{1}{2}\,\hat\psi(x)\left(\gamma^\mu\der_\mu-m\right)\psi(x),\label{dirac-lagrangian}\eea where the curved space gamma matrices are defined as usual: $\gamma^\mu=\gamma^a e^\mu_a$ ($e^a_\mu$ is the tetrad field obeying $g_{\mu\nu}=\eta_{ab}e^a_\mu e^b_\nu$, while $\gamma^a$ are the usual Dirac gamma matrices in the Minkowski space), $\hat\psi=\psi^\dag\gamma^0$ is the Dirac adjoint, and the sign of $m$ is irrelevant. In the above equations, for simplicity, I omitted the gauge terms like $i e A_\mu$, the spin connection, etc. In order to include these terms one should just replace the space-time derivative $\der_\mu$ by the covariant guage derivative $D_\mu=\der_\mu+ie A_\mu+ig W^i_\mu\sigma^i+...$, where the ellipsis stand for the spin connection terms. 

The stress-energy tensor for the Dirac spin $1/2$ field can be written as 

\bea T^{(\psi)}_{\mu\nu}=\hat\psi\gamma_\mu\der_\nu\psi-g_{\mu\nu}{\cal L}_\psi=\hat\psi\gamma_\mu\der_\nu\psi,\label{dirac-set}\eea where, in the last step, we took into account the fact that the spinor field equations are first order: $\left(\gamma^\mu\der_\mu-m\right)\psi(x)=0$, so that we can set ${\cal L}_\psi=0$. Notice that the signature reversal transformations: 

\bea m\rightarrow\pm im,\;g_{\mu\nu}\rightarrow-g_{\mu\nu},\label{sign-rev}\eea are a symmetry of the Dirac's equation only if, at the same time, $\gamma^\mu\rightarrow\pm i\gamma^\mu$, where the factor ``$i$'' may be ascribed to the transformation of the tetrad: $e_a^\mu\rightarrow\pm i e_a^\mu$ ($e^a_\mu\rightarrow\mp i e^a_\mu$), which induces the signature flip $g_{\mu\nu}\rightarrow-g_{\mu\nu}$.\footnote{Under this interpretation, the constant Dirac gamma matrices $\gamma^a$ and the constant Minkowski metric $\eta_{\mu\nu}$ are unchanged by the signature transformation \cite{duff}.} This implies that, under (\ref{sign-rev}), $\gamma^0\rightarrow\pm i\gamma^0$, while $\hat\psi\rightarrow\pm i\hat\psi$ and 

\bea &&\hat\psi\gamma^\mu\der_\mu\psi\rightarrow-\hat\psi\gamma^\mu\der_\mu\psi,\;m\hat\psi\psi\rightarrow-m\hat\psi\psi\nonumber\\
&&\;\;\;\;\;\;\;\;\;\;\;\;\;\;\;\;\;\;\;\;\;\;\;\;\Rightarrow\;{\cal L}_\psi\rightarrow-{\cal L}_\psi.\label{lagrangian-transf}\eea Besides, since $\gamma_\mu\rightarrow\mp i\gamma_\mu$, $\Rightarrow\;\hat\psi\gamma_\mu\rightarrow\hat\psi\gamma_\mu$, this means that under signature reversal the stress-energy tensor of the spin $1/2$ field (\ref{dirac-set}) is unchanged. This is more easily seen from the definition of the stress-energy tensor of matter:

\bea T_{\mu\nu}=-\frac{2}{\sqrt{|g|}}\frac{\der\left(\sqrt{|g|}{\cal L}\right)}{\der g^{\mu\nu}},\label{set-def}\eea which is unchanged by the simultaneous transformations $g_{\mu\nu}\rightarrow-g_{\mu\nu}$, ${\cal L}\rightarrow-{\cal L}$.

The following Einstein-Dirac equations can be derived from the action (\ref{gag-action}) with ${\cal L}(\psi,\der\psi)\equiv{\cal L}_\psi$ and ${\cal L}(\bar\psi,\der\bar\psi)\equiv{\cal L}_{\bar\psi}$ given by Eq. (\ref{dirac-lagrangian}):

\bea &&G_{\mu\nu}\equiv R_{\mu\nu}-\frac{1}{2}g_{\mu\nu}R=8\pi G\left[T^{(\psi)}_{\mu\nu}-T^{(\bar\psi)}_{\mu\nu}\right],\nonumber\\
&&\left(\gamma^\mu\der_\mu-m\right)\psi=0,\;\left(\gamma^\mu\der_\mu-m\right)\bar\psi=0,\nonumber\\
&&\der_\mu\hat\psi\gamma^\mu+m\hat\psi=0,\;\der_\mu\hat{\bar\psi}\gamma^\mu+m\hat{\bar\psi}=0,\label{dirac-einstein-feq}\eea where the stress-energy tensors for the gravitating Dirac field $T^{(\psi)}_{\mu\nu}$ and for the antigravitating one $T^{(\bar\psi)}_{\mu\nu}$, are given by Eq. (\ref{dirac-set}). Notice that the equations (\ref{dirac-einstein-feq}) are manifestly invariant under the G-aG transformations

\bea G\rightarrow-G,\;g_{\mu\nu}\rightarrow-g_{\mu\nu},\;\psi\rightarrow\bar\psi,\;\hat\psi\rightarrow\hat{\bar\psi},\label{gag-t}\eea which include the signature reversal transformation (\ref{sign-rev}). 

I want to underline that there is no problem with the unconventional negative sign of the momentum-energy tensor $T^{(\bar\psi)}_{\mu\nu}$ of the antigravitating spin $1/2$ Dirac field $\bar\psi$ in the Einstein equation in (\ref{dirac-einstein-feq}), since, as mentioned above, this sign can be absorbed into the Newton's constant $G$ which, for the antigravitating particles, is negative: $-G$, $$8\pi G\left[T^{(\psi)}_{\mu\nu}-T^{(\bar\psi)}_{\mu\nu}\right]=8\pi\left[GT^{(\psi)}_{\mu\nu}+(-G)T^{(\bar\psi)}_{\mu\nu}\right].$$ 

Implementing the above procedure in the case involving scalar fields is more difficult as we shall see. Nonetheless, scalar fields are central to the SMP in order to implement the acquirement of mass by the elementary particles and, if one wants to take for serious the G-aG symmetry, scalar fields can not be avoided.


\section{G-aG symmetry and scalar fields}\label{toy}

In this section I shall implement the approach of the former section in a toy model that is based on general relativity plus a pair of minimally coupled self interacting scalar fields. The Lagrangian for such a scalar fields is usually given by 

\bea {\cal L}_\chi=-\frac{1}{2}\left(\der\chi\right)^2-V(\chi),\label{scalar-f-lagrangian}\eea where the $\chi$ collectively stands for the gravitating and antigravitating partners $\phi$ and $\bar\phi$, respectively. As seen, due to the fact that under signature reversal the scalar field potential $V(\chi)$ is not transformed, while $(\der\chi)^2\rightarrow-(\der\chi)^2$, this Lagrangian does not transform under the G-aG transformations in the same way as the spin $1/2$ field Lagrangian (\ref{dirac-lagrangian}) does (see Eq. (\ref{lagrangian-transf})). 

Hence, in order to extend the approach of the former section to the scalar field sector, in the Lagrangian (\ref{scalar-f-lagrangian}) we introduce the following innocuous factor multiplying the potential $V$:

\bea \gamma\equiv\frac{G}{|G|},\;\gamma^2=1\;\Rightarrow\;\gamma=\pm 1,\label{gamma}\eea so that we end up with ${\cal L}_\chi=-(\der\chi)^2/2-\gamma V(\chi)$. Under the G-aG transformations: $G\rightarrow-G$ ($\gamma\rightarrow-\gamma$), $g_{\mu\nu}\rightarrow-g_{\mu\nu}$, this modified Lagrangian transforms as desired: ${\cal L}_\chi\rightarrow-{\cal L}_\chi$. The proposed action reads:\footnote{Notice I kept the same symbol $V$ for the self-interacting potential, meaning that the functional form of both $V(\phi)$ and $V(\bar\phi)$ is the same.}

\bea &&S=\int d^4x\sqrt{|g|}\left\{\frac{R}{16\pi G}-\frac{1}{2}(\der\phi)^2-\gamma V(\phi)\right.\nonumber\\
&&\left.\;\;\;\;\;\;\;\;\;\;\;\;\;\;\;\;\;\;\;\;\;\;\;\;\;\;\;\;\;\;\;\;\;\;\;\;\;\;+\frac{1}{2}(\der\bar\phi)^2+\gamma V(\bar\phi)\right\}.\label{master-action}\eea This action is manifestly invariant under the following G-aG transformations:

\bea G\rightarrow-G\;(\gamma\rightarrow-\gamma),\;g_{\mu\nu}\rightarrow-g_{\mu\nu},\;\phi\rightarrow\bar\phi.\label{gag-transf}\eea Besides if the potential $V$ is an even function, then the reflection $\phi\rightarrow -\phi$, $\bar\phi\rightarrow -\bar\phi$, is also a symmetry of (\ref{master-action}). 

The field equations derivable from the action (\ref{master-action}) are the following:

\bea &&G_{\mu\nu}\equiv R_{\mu\nu}-\frac{1}{2}g_{\mu\nu}R=8\pi G\left[T_{\mu\nu}^{(\phi)}-T_{\mu\nu}^{(\bar\phi)}\right],\label{einstein-feq}\\
&&\nabla^2\phi=\gamma\frac{dV(\phi)}{d\phi},\;\nabla^2\bar\phi=\gamma\frac{dV(\bar\phi)}{d\bar\phi},\label{kg-feq}\eea where $\nabla^2\equiv g^{\mu\nu}\nabla_\mu\nabla_\nu$ ($\nabla_\mu$ accounts for the covariant derivative operator). In Eq. (\ref{einstein-feq}) the stress energy tensor for the scalar fields $\phi$ and $\bar\phi$ is defined as usual (see the definition in Eq. (\ref{set-def})):

\bea T_{\mu\nu}^{(\chi)}=\der_\mu\chi\der_\nu\chi-\frac{1}{2}g_{\mu\nu}(\der\chi)^2-\gamma\,g_{\mu\nu}V(\chi),\label{set}\eea where, as before, $\chi$ is the collective name for the fields $\phi$ and $\bar\phi$. It is not difficult to check that the field equations (\ref{einstein-feq}), (\ref{kg-feq}), respect the G-aG symmetry (\ref{gag-transf}). 

Notice that the stress-energy tensor for the exotic scalar $T_{\mu\nu}^{(\bar\phi)}$ enters with the wrong sign in the right-hand side (RHS) of the Einstein's equation (\ref{einstein-feq}). From the physical standpoint this means that, if assume that $T^{(\bar\phi)}_{00}\geq 0$ is non-negative, the exotic field contributes with a negative effective energy: $\rho^\text{eff}_{\bar\phi}=-T^{(\bar\phi)}_{00}<0$, to the energy budget which sources the gravitational field. However, in view of the G-aG symmetry inherent in the theory (\ref{master-action}), (\ref{einstein-feq}), (\ref{kg-feq}), the ghost field $\bar\phi$ may be regarded as a standard scalar field with the positive energy density which, instead of gravitating, antigravitates. This feasible physical interpretation is apparent if one realizes that, in the RHS of the Einstein's equation (\ref{einstein-feq}), the minus sign in the second term may be absorbed into the Newton's constant: $8\pi[G\,T_{\mu\nu}^\phi+(-G) T_{\mu\nu}^{\bar\phi}]$. Hence, as it was for the fermion fields, once G-aG symmetry is adopted, for the scalar fields the negative energies are harmless in the present setup.


\section{Normalized gravitational charge}\label{g-charge}

In this section I want to make a few comments on the innocuous factor $\gamma$ defined in Eq. (\ref{gamma}) of the former section. Given the two possible eigenvalues of $\gamma$, it is evident from the definition that, for the gravitating fields $\chi$, the normalized gravitational charge $\gamma_\chi=+1$, while for the antigravitating fields $\bar\chi$, the normalized gravitational charge $\gamma_{\bar\chi}=-1$. In this vein the gravitational charge of a given field $\chi$ can be expressed as: $Q_\text{grav}=\gamma_\chi m^\chi_\text{grav}$, where $m^\chi_\text{grav}$ is the gravitational mass of the field.

At first sight it looks like if in Minkowski (nongravitational) background our G-aG symmetry, in the matter sector:\footnote{Since, as already explained, the sign flip of the gravitational metric $g_{\mu\nu}$ is due to the transformation of the tetrad: $e^a_\mu\rightarrow\mp i e^a_\mu$, hence, in the absence of gravitation, under the G-aG transformations the Minkowski metric is unchanged: $\eta_{\mu\nu}\rightarrow\eta_{\mu\nu}$.} $\eta_{\mu\nu}\rightarrow\eta_{\mu\nu}$, $\phi\rightarrow\bar\phi$, $\rho_\text{mat}\rightarrow-\rho_\text{mat}$, does not differ from the ``energy-parity'' symmetry \cite{sundrum}: $\eta_{\mu\nu}\rightarrow\eta_{\mu\nu}$, $\phi\rightarrow\bar\phi$, $H\rightarrow-H$ ($H$ is the matter Hamiltonian). Meanwhile, in the case when the gravitational interactions are turned on, our setup differs from that of Ref. \cite{sundrum} in that, while the gravitational action in (\ref{master-action}) respects the G-aG symmetry, it explicitly violates the energy-parity. 

In order to explain this apparent paradox let us show that, even in the absence of gravity, our setup clearly differs from the one in Ref. \cite{sundrum}. To make the differences evident in flat Minkowski space, lets go into the matter Hilbert space. Since we assume that the (normalized) gravitational charge $\gamma$ is conserved but, perhaps, for a very tiny violation in our present universe where gravity is switched on, then $\gamma$ commutes with the matter Hamiltonian $H$: $[H,\gamma]=0$. Actually, since $\gamma^2=1$ $\Rightarrow\;\gamma=\pm 1$, then $\left.\left.\gamma|\psi\right\rangle=\pm|\psi\right\rangle,$ and

\bea &&\left.\left.\left.H\gamma|\psi\right\rangle=\pm H|\psi\right\rangle=\pm E|\psi\right\rangle,\nonumber\\
&&\left.\left.\left.\gamma H|\psi\right\rangle=\gamma E|\psi\right\rangle=\pm E|\psi\right\rangle\;\Rightarrow\;[H,\gamma]=0,\nonumber\eea where the energy $E$ is an eigenvalue of the Hamiltonian $H$. Here I want to point out that, in correspondence with the two possible eigenvalues of the normalized gravitational charge $\gamma=\pm 1$, there are two possible eigenstates $\left.|\psi\right\rangle_E^\pm=\pm|\left.\psi\right\rangle_E$ with the same (positive) energy $E>0$. In Ref. \cite{sundrum}, on the contrary, it is postulated that the ``energy-parity'' symmetry operation $P$ ($P^2=1$), does actually anticommute with the Hamiltonian: $\left\{H,P\right\}=0$, so that, an energy eigenstate $\left.\left.H|\psi\right\rangle_E=E|\psi\right\rangle_E$ is transformed into one with the opposite energy $\left.\left.HP|\psi\right\rangle_E=-EP|\psi\right\rangle_E$ by the action of the energy-parity $P$. The fact that the operator $P$ does not commute with the Hamiltonian, means that, unlike the normalized gravitational charge $\gamma$, $P$ is not a conserved quantity, which means, in turn, that the energy-parity it is not an actual symmetry of (\ref{master-action}).


\section{Vacuum energy}\label{vacuum}

The introduction of a constant $\Lambda$ term in the action (\ref{gag-action}) breaks the G-aG symmetry. This fact hints at the possibility that, precisely, this kind of symmetry could be responsible for a zero value of the cosmological constant. Violations of this symmetry would yield to a net non-vanishing vacuum energy. Since models with spontaneous symmetry breaking are relevant to the cosmological constant problem \cite{ccp-sahni}, here I shall explore the model (\ref{master-action}) of the former section by considering symmetry breaking potetials $V(\phi)$ and $V(\bar\phi)$. A remarkable property of the model (\ref{master-action}) is that the Klein-Gordon equations (\ref{kg-feq}) for the fields $\phi$ and $\bar\phi$, coincide. The consequence is that both fields will tend to run down the potentials towards smaller energies. Therefore, if $V$ has global minima, both $\phi$ and $\bar\phi$ will tend to approach one of these minima. This is, precisely, the key ingredient in the present approach to explain the link between the G-aG symmetry and the vanishing of the vacuum energy. Actually, what matters to the effective vacuum energy

\bea \rho^\text{eff}_\text{vac}=V(\phi_\text{vac})-V(\bar\phi_\text{vac}),\label{vac-e}\eea is the difference in the self-interaction potentials at vacuum values of the fields $\left\langle\phi\right\rangle=\phi_\text{vac}$ and $\left\langle\bar\phi\right\rangle=\bar\phi_\text{vac}$, so that, if $V(\phi_\text{vac})=V(\bar\phi_\text{vac})$, the vacuum energy density vanishes. To illustrate this point, let us consider the ``Mexican hat'' potential \cite{rinaldi}:

\bea V(\chi)=\lambda\left(\chi^2-\frac{v^2}{2}\right)^2,\label{mexican-hat}\eea where $\lambda$ is the self-coupling, while $v$ is the symmetry breaking parameter ($v\simeq 246$ GeV). The symmetric state $(\phi,\bar\phi)=(0,0)$ is unstable and the system settles in one of the following ground states $(\phi_\text{vac},\bar\phi_\text{vac})$: $(v/\sqrt{2},v/\sqrt{2})$, $(v/\sqrt{2},-v/\sqrt{2})$, $(-v/\sqrt{2},v/\sqrt{2})$, $(-v/\sqrt{2},-v/\sqrt{2})$. This means that the reflection symmetry $\phi\rightarrow -\phi$, $\bar\phi\rightarrow -\bar\phi$, inherent in the theory (\ref{master-action}) with the potential (\ref{mexican-hat}), is spontaneously broken. Since $V(\phi_\text{vac})=V(\bar\phi_\text{vac})=0$, the immediate consequence is that the effective vacuum energy (\ref{vac-e}) vanishes: $\rho^\text{eff}_\text{vac}=0$. Hence, no net cosmological constant is left after reflection symmetry breaking, and the G-aG symmetry is preserved by the vacuum state. This result is not modified if, instead of (\ref{mexican-hat}), use the symmetry breaking potential $V(\chi)=\lambda\chi^4-2v^2\lambda\chi^2$ \cite{higgs}. In this latter case, at the minimums ($\chi=\pm v$): $V(\pm v)=-\lambda v^2$. Hence, $V(\phi_\text{vac})=V(\bar\phi_\text{vac})=-\lambda v^2$, so that no net vacuum energy (\ref{vac-e}) survives after symmetry breaking.  

At this point a comment is required: given that the initial conditions for the fields $\phi$, $\bar\phi$, in principle differ, it may arise that, while these fields are rolling down their potentials, a certain residual dynamical cosmological constant $$\Lambda(\phi,\bar\phi)=V(\phi)-V(\bar\phi)\neq 0\;\;(|\Lambda(\phi,\bar\phi)|\leq\lambda v^4/4),$$ might have inflated the universe. Yet the G-aG symmetry is preserved. As long as both fields settle down in the minimums of their potentials the residual cosmological constant (effective vacuum energy) vanishes as we explained above.

Which physical mechanism is actually responsible for a small violation of G-aG symmetry, is a question that could be clarified only once exotic antigravitating fields like $\bar\phi$, are built into a fundamental theory of the physical interactions, including gravity.\footnote{The existing standard model of the fundamental interactions, without the inclusion of gravity, is unable to differentiate gravitating and antigravitating particles.} In the absence of such fundamental theory of the unified interactions, one might only conjecture on
the physical origin of the G-aG asymmetry. In this regard, a possible origin of the aforementioned asymmetry can be explained using similar arguments than those used to explain the matter-antimatter asymmetry \cite{sakharov, m-am, bar-asym, bar-asym-1, pap-asym}, i. e., by invoking nonconservation of the gravitational charge. Since any initially generated G-aG asymmetry may be washed out by cosmological inflation, a postinflationary mechanism for generating the tiny asymmetry observed in the cosmological scales, is required. An alternative can be a mechanism of the kind in Ref. \cite{bar-asym-1}, which is based on the consideration of complex scalar fields and higher dimension operators that explicitly break the global $U(1)$ symmetry, where the necessary amount of asymmetry is generated, precisely, during inflation.


\section{G-aG symmetry and quintom models}\label{quintom}

There is an alternative to the introduction of the innocuous factor $\gamma=G/|G|$ of Eq. (\ref{gamma}), in the scalar fields self-interaction potentials in (\ref{master-action}). Actually a quintom-like Lagrangian model \cite{quintom}: $${\cal L}=\frac{R}{16\pi G}-\frac{1}{2}(\der\vphi)^2-V(\vphi)+\frac{1}{2}(\der\bar\vphi)^2-V(\bar\vphi),$$ respects the G-aG symmetry (\ref{gag-transf}). The field equations read:

\bea &&G_{\mu\nu}=8\pi G\left[T^{(\vphi)}_{\mu\nu}-T^{(\bar\vphi)}_{\mu\nu}\right],\nonumber\\
&&\nabla^2\vphi=\frac{dV(\vphi)}{d\vphi},\;\nabla^2\bar\vphi=-\frac{dV(\bar\vphi)}{d\bar\vphi},\label{phantom-feq}\eea where ${\cal L}_\vphi=-(\der\vphi)^2/2-V(\vphi)$ and ${\cal L}_{\bar\vphi}=-(\der\bar\vphi)^2/2+V(\bar\vphi)$, and the stress-energy tensors above are defined by (\ref{set-def}). This means that the scalar field Lagrangians, in this case, do not transform under signature flip in the same way as the former matter Lagrangians did. Nonetheless, under the simultaneous transformations: $g_{\mu\nu}\rightarrow-g_{\mu\nu}$, $\vphi\rightarrow\bar\vphi$, the difference ${\cal L}_\vphi-{\cal L}_{\bar\vphi}$ is not transformed, meanwhile $$T^{(\vphi)}_{\mu\nu}-T^{(\bar\vphi)}_{\mu\nu}\rightarrow-\left[T^{(\vphi)}_{\mu\nu}-T^{(\bar\vphi)}_{\mu\nu}\right].$$ Hence, the field equations (\ref{phantom-feq}) are invariant under the G-aG transformations: $$G\rightarrow-G,\;g_{\mu\nu}\rightarrow-g_{\mu\nu},\;\vphi\rightarrow\bar\vphi.$$

Unlike the model of (\ref{master-action}), the fields $\vphi$ and $\bar\vphi$ do not follow the same motion equations: while the scalar field $\vphi$ rolls down the potential $V(\vphi)$ towards (one of) the minimum(s), its antigravitating partner $\bar\vphi$ rolls up toward the local maximum of $V(\bar\vphi)$. In other words: $\bar\vphi$ rolls down towards the local minimum of the potential $-V(\bar\vphi)$. Assuming self-interaction potentials of the kind (\ref{mexican-hat}) and given convenient initial conditions, it may happen that in the vaccuum state $\vphi_\text{vac}=\pm v/\sqrt{2}$, while $\bar\vphi_\text{vac}=0$, so that a nonvanishing effective energy density is generated 

\bea &&\rho^\text{eff}_\text{vac}=V(\vphi_\text{vac})+V(\bar\vphi_\text{vac})\nonumber\\
&&\;\;\;\;\;\;\;=0+V(\bar\vphi_\text{vac})=\frac{\lambda}{4}\,v^4\sim 3.6\times 10^9\;\text{GeV}^4,\nonumber\eea leading to breakdown of the G-aG symmetry by the vaccuum state. Of course, this toy model which is sustained by potentials of the mexican-hat kind, produces an inadmissible large amount of breakdown of G-aG symmetry, which differs from the observed value by some 56 orders.

Anyway, the adoption of G-aG symmetry in quintom models means that the unconventional negative sign of the kinetic energy of the phantom field $\bar\vphi$ is harmless, since $\bar\vphi$ can be regarded as having conventional positive kinetic energy as long as it antigravitates, i. e., it feels a negative gravitational coupling $-G<0$. In this case, however, a cautionary note is to be made: although the antigravitating field $\bar\vphi$ has positive kinetic energy, its self-interaction potential can be a negative quantity, so that one has to care about the functional form of the potential in order to avoid any problem with the negative energies.


\section{Quantum instability}\label{instab}

Let us, first, to qualitatively discuss how the G-aG symmetry impacts the issue about the potential instabilities of the proposed scenarios when gravity is switched on. While negative energies may have catastrophic consequences in non-gravitational (in particular Minkowski) backgrounds, in backgrounds with a gravitational dynamics, thanks to the G-aG symmetry, the negative sign of the energy density may be harmless. Actually, as already shown, thanks to the interaction with gravity, which is quantified by the gravitational coupling $G$ (properly the Newton's constant), any negative energy density $\rho_\text{mat}<0$ in the RHS of the Einstein's equations (\ref{einstein-feq}), may be regarded as positive if, at the same time, the attractive character of the gravitational interactions is traded by gravitational repulsion (antigravitation) through the sign flip $G\rightarrow-G$. In consequence, the standard tenet that ``a negative-energy field can not have a (gravitating) vacuum state of minimal energy,'' i. e., that no consistent quantum field theory can be build out of such a field, may be wrong if adopt the G-aG symmetry as a fundamental principle of nature, since ``a vacuum state of minimal (positive) energy indeed exists which antigravitates.'' In this vein gravity-induced vacuum decay is prevented.

If assume a nongravitational background $g_{\mu\nu}=\eta_{\mu\nu}$, gravity-induced vacuum decay is surely avoided. However, in this case a more serious and problematic objection against negative Lagrangians/energies is related with potential instabilities originating from the direct matter couplings $\propto\phi^2\bar\phi^2$ between $\phi$ and $\bar\phi$, that might arise due to quantum effects. In general, these couplings would contribute to the decay of the vacuum.\footnote{As shown in \cite{sundrum} for the case when ``energy-parity'' symmetry is adopted, if assume these couplings to have their minimal natural length, they do not dominate any of the vacuum decay estimates and may be ignored. In other words, in Ref. \cite{sundrum} it has been shown that, in a cosmological context compatible with inflation and standard cosmology, the vacuum decay due to the coupling between the fields $\phi$ and $\bar\phi$ may be acceptably slow under reasonable assumptions.} 

In the following lines I shall discuss how the above picture with the apparently insurmountable instabilities, is modified by the adoption of G-aG symmetry. For this purpose let us explore the setup (\ref{master-action}) when the Minkowski space is considered. In this case the mentioned setup is represented by the effective Lagrangian: 

\bea {\cal L}\left[\phi,\bar\phi\right]={\cal L}\left[\phi\right]-{\cal L}\left[\bar\phi\right],\label{lag}\eea where, as customary, ${\cal L}\left[\chi\right]=-(\der\chi)^2/2-V(\chi).$ Actually, in the Minkowski space, under the G-aG transformations, the kinetic energy term $\propto(\der\chi)^2$ does not change sign.\footnote{Recall that the sign flip of the gravitational metric $g_{\mu\nu}$ is due to the transformation of the tetrad: $e^a_\mu\rightarrow\mp i e^a_\mu$. Hence, under $G\rightarrow-G$, $g_{\mu\nu}\rightarrow-g_{\mu\nu}$, the Minkowski metric is unchanged: $\eta_{\mu\nu}\rightarrow\eta_{\mu\nu}$. This is due to the trivial fact that in flat Minkowski space gravitation is absent, so that gravitation and antigravitation are indistinguishable phenomena.} This is why, in Eq. (\ref{lag}), I have removed the innocuous factor $\gamma$ multiplying the potentials. As a consequence, under the transformation: $\phi\rightarrow\bar\phi$, the Lagrangian (\ref{lag}) transforms like: ${\cal L}\left[\phi,\bar\phi\right]\rightarrow-{\cal L}\left[\phi,\bar\phi\right]$. Hence, in order to implement the approach of section \ref{toy} in the present case where gravity is switched off, I propose the following gravitationally weighted Lagrangian: 

\bea {\cal L}_\gamma\left[\phi,\bar\phi\right]=\gamma{\cal L}\left[\phi,\bar\phi\right]=\gamma{\cal L}\left[\phi\right]-\gamma{\cal L}\left[\bar\phi\right],\label{eff-lag}\eea which is invariant under the transformations:

\bea \gamma\rightarrow-\gamma,\;\eta_{\mu\nu}\rightarrow\eta_{\mu\nu},\;\phi\rightarrow\bar\phi.\label{transf}\eea 

In the absence of gravity Eq. (\ref{transf}) amounts to the G-aG transformations (\ref{gag-transf}). Due to the vanishing effects of gravity, in this case it is more convenient to write $\gamma=M^2_\text{Pl}/|M^2_\text{Pl}|$, so that the first transformation in (\ref{transf}) may be regarded as a transformation between fields with positive (normalized) gravitational charge $\gamma_\phi=+1$ ($M^2_\text{Pl}>0$), and their partners with the negative gravitational charge $\gamma_{\bar\phi}=-1$ ($M^2_\text{Pl}<0$). Hence, the ghost Lagrangian $${\cal L}_\text{ghost}=-{\cal L}\left[\bar\phi\right]=\gamma_{\bar\phi}\left[-\frac{1}{2}(\der\bar\phi)^2-V(\bar\phi)\right],$$ may be regarded, alternatively, as a gravitationally weighted Lagrangian for a field with the positive energy, but with negative normalized gravitational charge $\gamma_{\bar\phi}=-1$. 

An alternative explanation of how the multiplication of the Lagrangian by the innocuous factor $\gamma$ changes the whole picture, can be given if go into the Hilbert space of states $\left.|\psi\right\rangle$, and consider the Hamiltonian\footnote{Here the dot means derivative with respect to the time coordinate, while $\nabla$ stands for the 3-dimensional divergence.} 

\bea H=\int d^3x\,{\cal H},\;{\cal H}=\frac{\dot{\bar\phi}^2}{2}+\frac{1}{2}(\nabla\bar\phi)^2+V(\bar\phi).\label{hamilton}\eea 

As already explained in section \ref{g-charge}, given that the normalized gravitational charge $\gamma=\pm 1$ commutes with the Hamiltonian, the negative energy eigenstate $\left.H|\psi\right\rangle_{-E}=-\left.E|\psi\right\rangle_{-E}$, may be regarded as a positive energy eigenstate $\left.|\psi\right\rangle^-_E$ of the gravitationally weighted (self-adjoint) Hamiltonian operator: $H_\gamma\equiv\gamma H=H\gamma=H^\dag_\gamma$, corresponding to the negative eigenvalue of the gravitational charge operator. Actually, since $$\gamma\left.|\psi\right\rangle^-=-\left.|\psi\right\rangle^-,\;\left.H|\psi\right\rangle_E=\left.E|\psi\right\rangle_E,$$ then $$H_\gamma\left.|\psi\right\rangle^-_E=H\gamma\left.|\psi\right\rangle^-_E=-\left.E|\psi\right\rangle^-_E,$$ where, as said, $\left.|\psi\right\rangle^-_E$ is an eigenstate of the gravitationally weighted Hamiltonian $H_\gamma$ characterized by positive energy $E>0$ and negative normalized gravitational charge $\gamma_\psi=-1$. Hence this state is repelled by the gravitating matter and, in order to excite it, positive energy $\geq E$ is required.

By the same procedure explained above, in the case of the quintom models (see section \ref{quintom}), one has (no gravity) 

\bea &&\gamma\left[-\frac{1}{2}(\der\vphi)^2-V(\vphi)\right]+\gamma\left[\frac{1}{2}(\der\bar\vphi)^2-V(\bar\vphi)\right]=\nonumber\\
&&\gamma\left[-\frac{1}{2}(\der\vphi)^2-V(\vphi)\right]-\gamma\left[-\frac{1}{2}(\der\bar\vphi)^2+V(\bar\vphi)\right].\nonumber\eea In this case the phantom Lagrangian (second term in the last line above) is traded by the negative gravity-charged ($\gamma_{\bar\vphi}=-1$) Lagrangian of a tachyon field with negative mass squared $-m^2=-\der^2V/\der\bar\vphi^2<0$, but with positive kinetic energy density $-(\der\bar\vphi)^2/2>0$. In consequence, I have transferred the non-unitarity/absence of a stable vacuum state problem, which is associated with ghosts, to instabilities associated with propagating tachyonic fields. Fortunately, for the usual tachyonic instability which arises in a Lorentz invariant theory when some field has a negative mass squared $m^2<0$, the time-scale $\tau$ of the instability is set by the inverse mass, $\tau\sim 1/|m|$, which may be very long compared to the other characteristic time scales if $m^2$ is small enough \cite{tachyons}, i. e., conventional tachyonic instabilities are present only for long wavelength modes, so that their rate of growth is bounded. As a consequence the instabilities associated with the phantom field \cite{phantom} in quintom models are much more harmful and less controlled than those due to their tachyonic alternative.


\section{Conclusion}

Although the naive approach explored in this paper is classical and far from realistic -- recall that the toy models of sections \ref{toy} and \ref{quintom} include only a couple of scalar fields so that these should be complemented with the inclusion of other boson and fermion fields of the SMP -- and, besides, it is incomplete in that it gives no explanation about a feasible mechanism for the present small violation of G-aG symmetry, nevertheless, it gives valuable insight on a possible connection between this (would be) fundamental symmetry of nature and the vanishing of the vacuum energy density. At the same time, on the light of the G-aG symmetry, the occurrence of negative energies in general relativity is much less dangerous than it seems. This is at the cost, however, of doubling the number of fields of the SMP, a process that bears no physical consequences, unless the gravitational interactions are switched on. In this latter case the relative strength of the gravity-antigravity effects, is suppressed by a factor of $10^{-36}$, when compared with the strength of the electromagnetic effects set by $\alpha\approx 1/137$. Otherwise, since only at energy scales of the order of the Planck energy $\sim 10^{19}$ GeV, are the effects of gravity comparable to those of the remaining interactions, it will very difficult to observe any effects coming from the G-aG symmetry, or from its absence, in current accelerator experiments. Perhaps cosmology will be a better suited arena where to look for signatures of the ``would be'' symmetry between gravity and antigravity.

Thinking in a cosmological scenario, one can guess that the starting state of the universe was one in which there was an exact balance between gravitating and antigravitating particles (no net gravity effects). Then, due to some initial density perturbations, gravitating particles progressively attracted among then and repelled antigravitating particles and so on, so that both kinds of fields nucleated in separate spacetime domains with growing volume. This marked the beginning of the cosmological expansion. I suspect that this kind of separate nucleation of gravitational and antigravitational domains occurred everywhere so that our particular universe evolved within one such domain. Hence, if the present proposal were correct, one would expect that somewhere out there antigravity drives the destiny of (perhaps very large) parts of the universe, in a similar way as gravity dictates the destiny of our particular universe. All of this is, of course, pure speculation.

If take seriously the possibility to have antigravitating partners of each particle of the standard model, several interesting questions should be addressed: Is the total gravitational charge of the universe positive, negative or exactly zero? In other words, is the universe as a whole G-aG symmetric while locally the G-aG symmetry is not satisfied? or, is it dominated by gravitating or antigravitating objects? Another interesting question is related with the kind of antigravitating objects that may exist in nature. While a satisfactory answer to these -- and perhaps other deep questions -- has to wait for more profound investigation of the physical consequences of complementing the SMP with antigravitating particles, the idea seems exiting and we are invited to explore its many faces.

\begin{acknowledgments}

I am grateful to Aharon Davidson for encouraging comments on an earlier version of this paper (Ref. \cite{quiros-2004}). I also thanks Massimiliano Rinaldi for useful comments and the SNI of M\'exico for financial support of this research.

\end{acknowledgments}




\end{document}